\documentclass[%
 reprint,
 superscriptaddress,
 amsmath,amssymb,
 aps,
pra,
floatfix,
]{revtex4-1}
\usepackage{xcolor}
\usepackage[final]{graphicx}
\usepackage{dcolumn}
\usepackage{bm}
\usepackage{systeme}
\usepackage{amsmath}
\usepackage{mdframed}
\usepackage{float}
\usepackage[caption = false]{subfig}
\usepackage{hyperref}

\begin{document}


\title{Social dilemma in traffic with heterogeneous drivers}


\author{Ricardo Sim\~ao}%
 \email{summernightdream@fis.grad.ufmg.br}
\author{Lucas Wardil}
\email{wardil@fisica.ufmg.br}
 \affiliation{Departamento de F\'\i sica, Universidade Federal de Minas Gerais,
Caixa Postal 702, CEP 30161-970, Belo Horizonte - MG, Brazil.
}%




\date{\today}

\begin{abstract}

There is a ``tragedy of the traffic'' analogous to the ``tragedy of the commons'' that can be caused by overtaking. We analyze the effect of overtaking in  a minimal model of vehicular traffic, the Nagel-Schreckenberg model, with two types of drivers: drivers that overtake and drivers that do not. We show that, under certain circumstances, overtaking is good because it increases the road capacity and minimizes the driver's mean time spent on the road. However, when these conditions are not met, overtaking is harmful to all. More specifically, we found that a social dilemma emerges in the vicinity of the transition to the congested traffic if the probability of random deceleration is low, which can also happen in more realistic single-lane models. The essential mechanism creating the social dilemma is the abrupt deceleration when the overtaking car returns to its lane. We analyze how the payoffs depend on the frequency of strategies in the population to conclude that the drivers that overtake are defectors and the ones that do not are cooperators, analogous to the strategies in ``tragedy of the commons'' class of games.


\end{abstract}

\maketitle


\section{Introduction}

Overtaking on single-lane roads can be daunting even for experienced drivers. Whenever possible, drivers in a hurry would like to overtake the slow ones and everyone would be satisfied, as the slow drivers would keep their pace and the fast ones would save time. However, transportation systems exhibit many unexpected social phenomena related to the tragedy of the commons \cite{tanimoto2018evolutionary}. The individuals want to reach their destinations safely and as fast as possible. However, the limited availability of space may create social dilemmas. A trivial example that clearly shows the struggle for a common resource is the choice between using public transportation or private vehicle at moderate or high traffic densities. Because public transportation is generally slow, the drivers may arrive at their destination faster if they use their cars instead of public transportation. However, jammed traffic sets in because of the large number of vehicles and, as a result, all individuals may spend more time in the traffic. Another example is the increase of traffic when an additional fast highway is built to connect two previously unconnected locations, known as the Braess' paradox \cite{hagstrom2001characterizing}. The temptation to take the fast lane to reduce the time travel may attract all drives to the fast highway, creating congestion that otherwise would never happen \cite{karlin2017game}.

The long commutes are becoming an increasing problem in the big metropolis. People experiencing them are more stressed, with serious implications for their well-being \cite{longcommute}. It is a natural assumptions that drivers would like to overtake slow vehicles if it reduces their commute times. The most basic environment which allows overtaking is the conventional two-way road with one lane in each direction. The drivers should abide by the traffic rules, driving along their lane and overtaking only when it is allowed and safe. Empirical studies on overtaking in two-way highways are scarce. Typically, these studies measure the headway distance when the overtaking starts, the length and the duration of the overtaking, the speed differences, the distance of moving back to lane \cite{passing2}. Three types of overtaking strategies have been observed in these environments: the flying, whereby the driver overtakes without reducing the speed; the acceleration, whereby the driver first reduces the speed and, when allowed, accelerates to overtake; and the ``piggybacking'', whereby the driver follows another driver that is overtaking \cite{passing1,passing2}.  These three strategies are considered safe strategies. The safety conditions are determined essentially by the gap availability in front of the slow car, the gap in the lane with the opposite flow, and the relative speeds. However, drivers can underestimate the safety criteria or can even become impatient, performing dangerous manoeuvres \cite{impatient}. In particular, the proximity of the oncoming vehicle can force the overtaking driver to move back to the lane as soon as possible. In one study, it was found that in $10\%$ of the cases the drivers were forced to move back to their lane \cite{Mocsari_1}.

Vehicular traffic is a very complex system. Important developments have been made at the theoretical level through computer simulations, empirical data analysis and observation of human experience in simulation machines (see the excellent reviews \cite{nagatani2002physics,chowdhury2000statistical,kerner2012physics}). The Nagel and Schreckenberg's cellular automaton model (NaSch model) is a well studied minimal model of vehicular traffic in single-lane environments \cite{nagel1992cellular}. This model is important because, despite its simplicity, it reproduces fundamental features of vehicular traffic such as the transition from the free-flow traffic to the congested traffic and the spontaneous formation of jams. The NaSch model is closely related to the asymmetric simple exclusion process \cite{lighthill1955kinematic,nagatani2002physics} and to stochastic growth models of one-dimensional surfaces in a two-dimensional medium \cite{barabasi1995fractal,chowdhury2000statistical}. Many generalizations of the NaSch model have been done and. A few examples are the addition of a slow-to-start rule \cite{takayasu19931}, multi-lane non-homogeneous environment \cite{moussa2010simulation},  probability of random deceleration depending on the velocity of the car \cite{barlovic1998metastable,iannini2017traffic}, inclusion of brake light to produce a more comfortable driving experience \cite{realistic} and non-null probability of accidents \cite{boccara1997car,moussa2003car,bentaleb2014simulation}. More examples can be found in the reviews \cite{chowdhury2000statistical, nagatani2002physics}. Comparison between the models and the empirical data  can be found in \cite{neubert1999single,kerner1996experimental,knospe2004empirical}. A specially important example for our discussion is a stochastic overtaking strategy implemented in the NaSch model in \cite{su2016effects}. In their work, when some conditions are satisfied, the overtaking happens with probability $q$. The authors show that increasing $q$ also increases  traffic flow. They also comment that when the probability of overtaking is high, the system remains in a high-flow regime even for high concentrations of cars \cite{su2016effects,su2016occurrence,su2018mean} due to an over-acceleration of the overtaking vehicles implicit in their algorithm. However, overtaking should play a negligible role in the high concentration traffic because there is little space available to the manoeuvre.

The possibility of overtaking may put the  individuals' self-interest in conflict with the interest of the others, giving rise to social dilemmas. A social dilemma is characterized by the possibility of choosing a strategy, the defection strategy, that provides higher payoffs to the self at the expense of the others, but making the population worse off if all adopt the defection strategy instead of  the cooperative strategy \cite{dawes1980social}. In one work, the authors implemented a model with two-lanes and introduced two types of drivers \cite{tanimoto2014dangerous,tanimoto2016social}. The cooperators are stuck in the slow lane, whereas the defectors may change lanes. The authors show that, if the car density is high, a weak social dilemma emerges because of the perturbations caused by the vehicles in the fast lane when fluctuations in the local distribution of vehicles appear. 

Here, we study the potential conflict between self-interest and social payoff that may arise as the result of overtaking in the rather simple scenario of the single lane. Overtaking in single-lane models can be interpreted as an approximation to overtaking in undivided two-way road in the limit of negligible opposite flow, where the drivers can use the free opposite lane to overtake. This situation is typical of the  morning and late commutes, where drivers go to downtown in the morning and return home in the evening. Even if there is a non-negligible opposite flow, overtaking in single-lane can model the behavior of risk-prone drivers that change lanes quickly.  In contrast to the model analyzed in \cite{su2016effects,su2016occurrence,su2018mean}, which analyzes a homogeneous population where all drivers adopt the same a stochastic strategy,  we introduce two types of drivers: cooperators, that follows the NaSch model, and defectors, that try to overtake whenever possible.  We show that overtaking can create a social dilemma  at intermediate concentration of vehicles if the probability of random deceleration is small. We  discuss the nature of the interactions and the proper way to quantify the payoff of the individual. The classification as defectors and cooperators is closely related to the usual classification used in the Tragedy of the Commons class of game. Finally, we analyzed more real models that soften the acceleration of the vehicles and we found out that the drastic deceleration occurring when a vehicle completes the overtaking is the essential mechanism creating the social dilemma.

This paper is organized as follows. We describe the model in Sec. \textbf{II} (the details of the algorithm are placed in the appendix). The main results are split into three parts: in the Sec. \textbf{III.A}, we show the analytical results; in the Sec. \textbf{III.B} we introduce the main the concepts and we analyze the social dilemma;  in  the Sec. \textbf{III.C} we implement the overtaking strategy to a more realistic model, known for its accuracy to experimental data regarding single-lane traffic, and discuss the mechanisms responsible for the emergence of the dilemma. We finish in the Sec. \textbf{IV} with a brief discussion of the results and the conclusion.

\section{The model}

The NaSch model represents vehicular traffic as a discrete process in space and time. The cars are represented as particles moving on a 1-dimensional lattice with periodic boundaries.  Each site is either occupied by a single car or empty, and the velocities of the vehicles assume values in the set $\{0,1,2,\ldots,5\}$. The vehicles following the NaSch dynamics (henceforth called \textit{cooperators}) update their position and velocity in four steps as follows. Let $x_n$ and $v_n$ be the position and the velocity of the $n$-th car at time $t$. In the first step, the speed is increased in one unit: $ v_n \rightarrow \min(v_{n}+1,v_{max})$. In the second step, the speed may be reduced  to avoid collision with the front car: $v_n  \rightarrow \min(v_n,(x_{n+1}-x_n)-1)$. In the third step, there is a chance of a randomly deceleration, which happens with probability $p$: $v_n  \rightarrow \max(v_n-1,0)$. After all velocities are updated, the position is updated: $\vec{x}_n \rightarrow \vec{x}_n+\vec{v}_n$.  We should mention that if the random deceleration comes before the second step, it would have a null effect whenever the second step takes effect. The random deceleration is necessary to scale the negative effects caused by the deceleration in the second step, which initiates the  spontaneous formation of jams  \cite{chowdhury2000statistical,nagatani2002physics,kerner2012physics}.

Based on the behavior of some drivers that use the lane with the opposite flow to make the overtaking manoeuvre, which is observed in experimental data \cite{passing1,passing2}, we made a minimal change to the NaSch model to allow overtaking when safety conditions are met. We do not consider any constraints coming from the opposite lane traffic, which is assumed to be negligible in our analysis. The dynamics is, therefore, effectivelly  unidimensional  with small windows exhibiting non-unidimensional dynamics when overtaking happens (for the necessity of such windows for the overtaking see \cite{nagatani2002physics}). The update rule for the vehicles following our algorithm (henceforth called \textit{defectors}) is very similar to the NaSch model. The defectors will attempt to overtake as many vehicles as possible, constrained to the condition that the defector will overtake only if its velocity is high enough to arrive in front of the overtaking vehicles considering their updated speeds. If it is not possible to overtake the last vehicle within its range, the defectors will try to overtake one less vehicle, until overtaking is no longer possible. In this case, the defector behaves as a cooperator. Since the random deceleration factor may be interpreted as the driver's response to the imperfections of the highway \cite{bouadi2016effect}, and both strategies coexist in the same environment, both defectors and cooperators are subjected to the same random deceleration. Figure \ref{fig0} illustrates the update rule of the defectors. See the Appendix for the details of the algorithm.

\begin{figure}[tbh]
\includegraphics[scale=0.75]{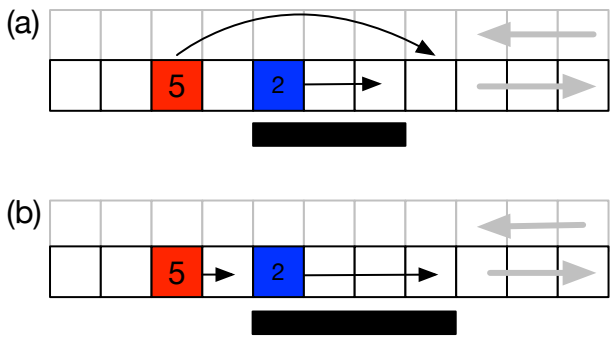}
\caption{Illustration of the overtaking rule. The blue and the red squares represent a driver to be overtaken and a vehicle trying to overtake, respectively. The numbers inside the squares are their current speed, the black rod represents the restriction imposed by the safety condition, and the arrows indicate the position in the next step. In this example, the defector does not randomly decelerate. In panel (a), the cooperator accelerates one unit and decelerate one unit due to the random deceleration.  Because the defector can land in a site after the cooperator, the overtaking is successful. Notice that in the next step the blue vehicle will have a velocity equal to one, whereas the red vehicle may keep its maximum velocity. In panel (b), the cooperator accelerates one unit and does not decelerate, advancing three sites.  Because the defector does not have enough velocity to arrive after the black rod, he must reduce his velocity to one, behaving effectively as a cooperator.}
\label{fig0}
\end{figure}


\section{Results}


To analyze the effects of individual behavior in the traffic and to understand the incentive structure at the individual level, we focus on two measures: the flux of vehicles and the individual payoff. 
The \textit{flux of vehicles} is the number of vehicles crossing a detector per time unit \cite{chowdhury2000statistical,kerner2012physics}. The fundamental hypothesis that at any point of the road the flow is a function of the density of cars suggests that the flux versus concentration diagram, also known as the fundamental diagram, characterizes the system \cite{lighthill1955kinematic}. Although this hypothesis may not be consistent with all traffic phenomena \cite{kerner2012physics}, the fundamental diagram is a good descriptor of the average macroscopic properties of the system. The average flux, $\rho$, is calculated as $\rho=\bar{v}c$, where $\bar{v}$ is the average velocity and $c$ is the global density of cars.

The \textit{individual payoff} takes into account all the factors that influence the decision of one individual. In our model, the core of the payoff is the commute time. Since the size of the road is  constant and no alternative routes are available, the only way to minimize the commute time is to drive faster. Hence, the average velocity of the individual is a natural measure of the individual payoff. Note that the average flux is a measure of the road capacity, which is fundamentally distinct from the concept of individual payoff. The free-flow regime is a good example to illustrate the distinction between average flux and individual payoff.  In the free-flow regime, the cars are most of the time at their maximum velocity. Thus, it is not possible to increase the individuals' payoff. However,  if the density increases (the system being still in the free-flow regime), the average velocity remains the same and the flux increases. Therefore, the average flux is not the proper quantifier to measure neither the individual's nor the social payoffs, the latter being considered as the average individual payoff. 

Thus, the following questions arise:  is overtaking good for the population as a whole? Is there any social dilemma? If so, which mechanism creates the social dilemma?  In the next sections, we provide answers to these questions.

\subsection{Insights based on the analytical results}

\begin{figure}[tbh]
\includegraphics[scale=.51]{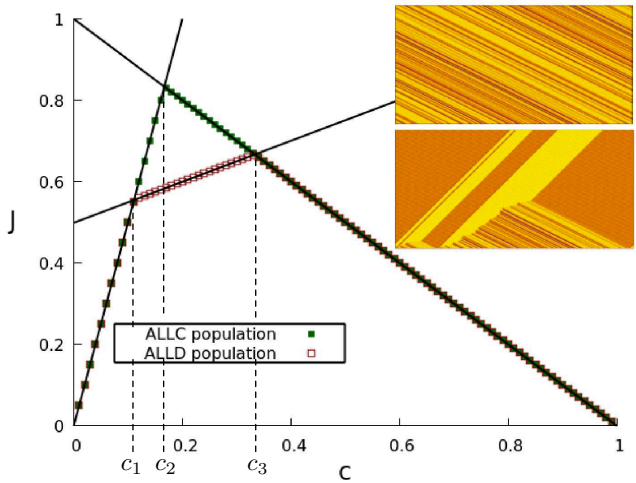}
\caption{ Fundamental diagram for the homogeneous populations of cooperators and defectors. The lines represent the functions $J(c)=5c$ , $J(c)=.504c + .499$, and $J(c)=1-c$ deduced in the text.  The images in details are the temporal-spatial pattern of the ALLC population (above) and ALLD population (below), both initialized under the same random spatial configuration and velocities at $p=0$ and $c=0.2174$. The darker the region the denser the local configuration. Time is measured from below to top and space from right to left. Notice that the 2-cycle emerges in the transient regime and absorb all the non-homogeneities of the system increasing in size. The values of $c_1$, $c_2$, and $c_3$ are provided in the main text.}
\label{figp0}
\end{figure}

The deterministic model, $p=0$, allows  analytical treatment, shedding light to the nature of our problem. The fundamental diagram for a homogenous population of cooperators (ALLC) and for a homogenous population of defectors (ALLD) with $p=0$ is shown in Fig. \ref{figp0}. Notice that at intermediate values of concentration the average flux of a population of defectors is lower than that of a population of cooperators. In homogeneous populations, the average flux is easily related to the average individual payoff at constant density by $J=c\bar{v}$. Thus, the average individual payoff in ALLC population is higher than that in the ALLD population. In other words, overtaking can be detrimental to the population.

To arrive at the analytical expression of the flux, let us begin with the ALLC population, where the individuals follows the NaSch model. At low values of $c$, the vehicles reach their maximum velocity for any initial condition (after a short transient interval) and are at least $v_{max}$ sites apart. As more vehicles are added to the system, the mean distance between the vehicles is reduced and, at $c_{2}=1/(1+v_{max})$, all  vehicles are exactly $v_{max}$ sites apart. Thus, the flux of the ALLC population for $c<c_{2}$ is given by the function 
$$J(c)=v_{max}c.$$ 
To analyze the case  $c>c_{2}$, let us define the total velocity in the stationary state, $\mathcal{V}$, as the sum of the velocities of all vehicles, that is, $ \mathcal{V}= \sum_{i=1}^{N}v_{i}$. Is is easy to see that for $c<c_{2}$ we have that $\mathcal{V}=Nv_{max}$, where $N$ is the number of vehicles. In the congested phase, the velocities of the vehicles depend on the space available. In a stationary configuration (stationary except for translations), the velocities correlate strongly with the available space, and $v_{i}$ can be approximated by $v_{i}=x_{i+1}-x_{i}-1$. Then,  
$$ \mathcal{V}= \sum_{i=1}^{N}v_{i} = \sum_{i=1}^{N} x_{i+1}-x_{i}-1 = L-N.$$
Notice that we have used the boundary condition $x_{i+N}=x_{i}+L$. Thus, the analytical expression of the flux in the congested phase is given by  
$$J(c)=c\bar{v}=(N/L)(\mathcal{V}/N) = \mathcal{V}/L=1-c.$$  
The expression of the flux for $c>c_{2}$  does not depend on the maximum velocity $v_{max}$ and is a good approximation for any maximum velocity. Also, because $v_{max}$ appears explicitly at the expression of $c_2$, there is a nice geometrical visualization: the perpendicular which crosses the $J(c)$ curve at the critical density $c_{2}$ forms an isosceles triangle with the $c$ axes, having the congested phase line as the hypotenuse for any value of $v_{max}$. 


The behavior of the ALLD population is identical to the behavior of the ALLC population at low concentration. Thus, the flux is given by $J(c)=v_{max}c$. However, at a concentration $c_1$ ($c_1<c_2$) an interesting feature emerges. Let $\#$ represent an empty site and suppose that the configuration $2\#\#20\#\#$ appears (the number represents the velocity of the vehicle localized in that site). Following our algorithm, this configuration evolves to $\#\#2\#\#13$. Notice that the $\#13\#$ is a forbidden configuration in the NaSch model and is often referred to as  Garden of Eden (GoE) state \cite{schadschneider1999nagel,hauert2006}. This configuration evolves to $\#2\#\#20\#$, which is identical to the initial one, but translated in one unity. Thus, there is a dynamical 2-cycle moving forward with mean velocity given by $v_{cy}=0.5$. If the local region is dense enough, there will be an increasing accumulation of vehicles behind the 2-cycle  (the $\#\#2\#\#2$ pattern). Since the 2-cycle loses one vehicle each two time steps, if the environment feeds the leftmost part of the pattern at a higher rate, the pattern will grow. The critical rate is one vehicle each two time steps. Assuming that the vehicles feeding the pattern are in the free state, then in two time steps they move $2v_{max}$ sites. Meanwhile, the pattern moves one site forward. Remembering that the fundamental unity of the pattern is $2\#\#$ (a vehicle effectively occupy three sites), this implies that at a concentration of $c_{1}=1/(2v_{max}-1)$ (notice that $c_{1}<c_{2}$) the rate of evaporation of vehicles will be the same as the rate of absorption. Consequently, the dynamics of these patterns will be stable. If $c > c_{2}$, this phase will emerge during the transient state and grow, except in initial eigen-configurations of the NaSch algorithm which do not contain the  local configuration $20\#\#$ anywhere. The eigen-configurations  have vanishing small probabilities at the thermodynamic limit if random initial configurations are chosen and, therefore, the NaSch like pattern loses stability to the emerging 2-cycle. It turns out that the  size of the 2-cycle pattern depends on the local density of vehicles and is as a self-organizing mechanism. The pattern will grow where there are many particles and shrink where there are few. The result is the coexistence of two highly organized phases: a free phase with density proportional to  the evaporation rate of the 2-cycle pattern ($c=1/(2v_{max}-1)$), and mean velocity given by $\bar{v}_f=v_{max}$; and the organized 2-cycle pattern with mean velocity of $v_s=2$ and density $c=1/3$. 
 Increasing the density above $c_1$ will increase the 2-cycle pattern size (its evaporation rate is constant) until there is no more space available, which happens at a global density equivalent to its own internal density $c_{3} = 1/3$. We may approximate the flux of vehicles summing the flux of the free phase to the flux of the slower 2-cycle pattern pondered by their characteristic density at the global density $c$: 
\begin{equation*}
J(c)=[1+(c-c_{1})/(c_{1}-c_{3})]c_{1}v_{max}+[(c_{1}-c)/(c_{1}-c_{3})]c_{3}\bar{v}_{s}.
\end{equation*}
For our parameters we have that $J(c) \approx 0.504c + 0.499$.   At concentrations higher than $c_3$, the cycle is destroyed by the perturbations caused by the extra vehicles and, as overtaking is no longer possible due to the high symmetry produced by the slow pattern, the Nasch-like behavior reemerge,  and $J(c)=1-c$.

To sum up, the flux difference between the ALLD and the ALLC population is then given by
\[ \Delta J(c)=\begin{cases} 
      0 & c\leq c_{1} \\
      f(c)-v_{max}c & c_{1}\leq c\leq c_{2} \\
      f(c)-(1-c) & c_{2}\leq c\leq c_{3} \\
      0 & c \geq c_{3}
   \end{cases}
\]
where $f(c)$ is the flux in the ALLD population $c_{1}\leq c\leq c_{3}$. Because the flux difference  is negative in the interval $c_{1}\leq c \leq c_{3}$ and $\bar{v}=J/c$, the deterministic case shows that overtaking can indeed be detrimental for everyone.

\subsection{Simulation results for $p\ne0$} 

\begin{figure*}[tbh]
\includegraphics[scale=0.47]{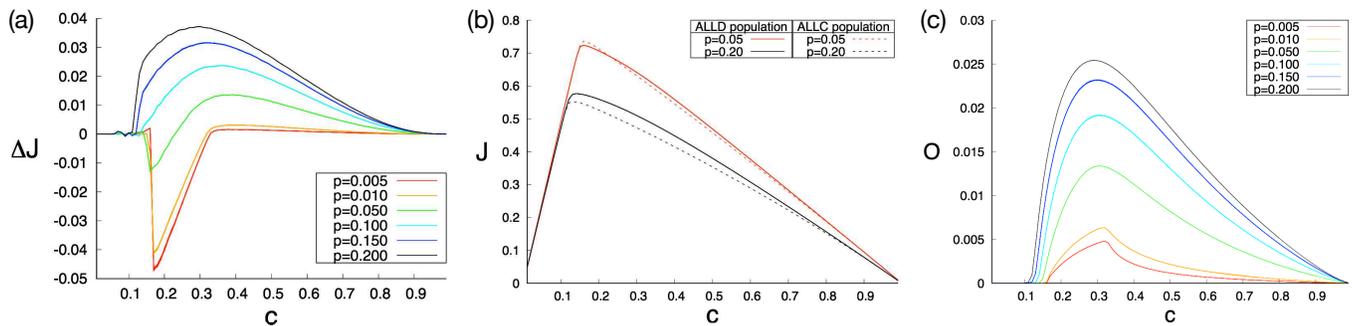}
\caption{Fundamental diagram analysis. (a) The fundamental diagram of homogeneous populations for different values of the probability of random deceleration.The diagram has two typical behaviors: free-flow and congested traffic.  If $p$ is sufficiently small, there is a region near the transition where the ALLD curve lies beneath the ALLC diagram. At high $p$, the ALLC curve lies entirely under the ALLD curve.  Also, notice that the curves for both populations are nearly the same at the free-flow regime and converge to the same values as the density gets higher. The standard deviation is smaller than the size of the symbols. (b) The difference of the flux between the homogeneous populations of cooperators and defectors for many $p$ values. The difference is given by  $\Delta J=J_d-J_c$, where $J_d$ and $J_c$  are the fluxes in each homogeneous populations. (c) The mean number of overtaking per vehicle per time unit, $O$, in function of the global concentration $c$. Notice that the highest number of overtaking happens near the onset of jammed behavior because of the large variance of the velocities and the availability of space for a safe overtaking.}
\label{fundamental-diagram}
\end{figure*}

The deterministic limit does not include random deceleration, which is a very important component  of the NaSch  model. In this session, we investigate with computer simulations the behavior of the system when $p\ne 0$. Figure \ref{fundamental-diagram} shows the fundamental diagram of the ALLC and ALLD populations for different values of $p$. 
The difference between the ALLC and the ALLD fluxes is very small at low and high concentrations. The reason is that overtaking is improbable at low concentrations due to the high velocity of the vehicles and, at high concentrations, due to the lack of free space. However, near the transition from the free to the congested traffic, the overtaking behavior shows up and a non-trivial behavior dependent on the probability of random deceleration emerges: for low values of $p$ the average flux of the ALLC population is higher than that of the ALLD population. Thus, there is a conflict between the individual self-interest and the common good under these conditions. 

\begin{figure}[h!]
\includegraphics[scale=0.45]{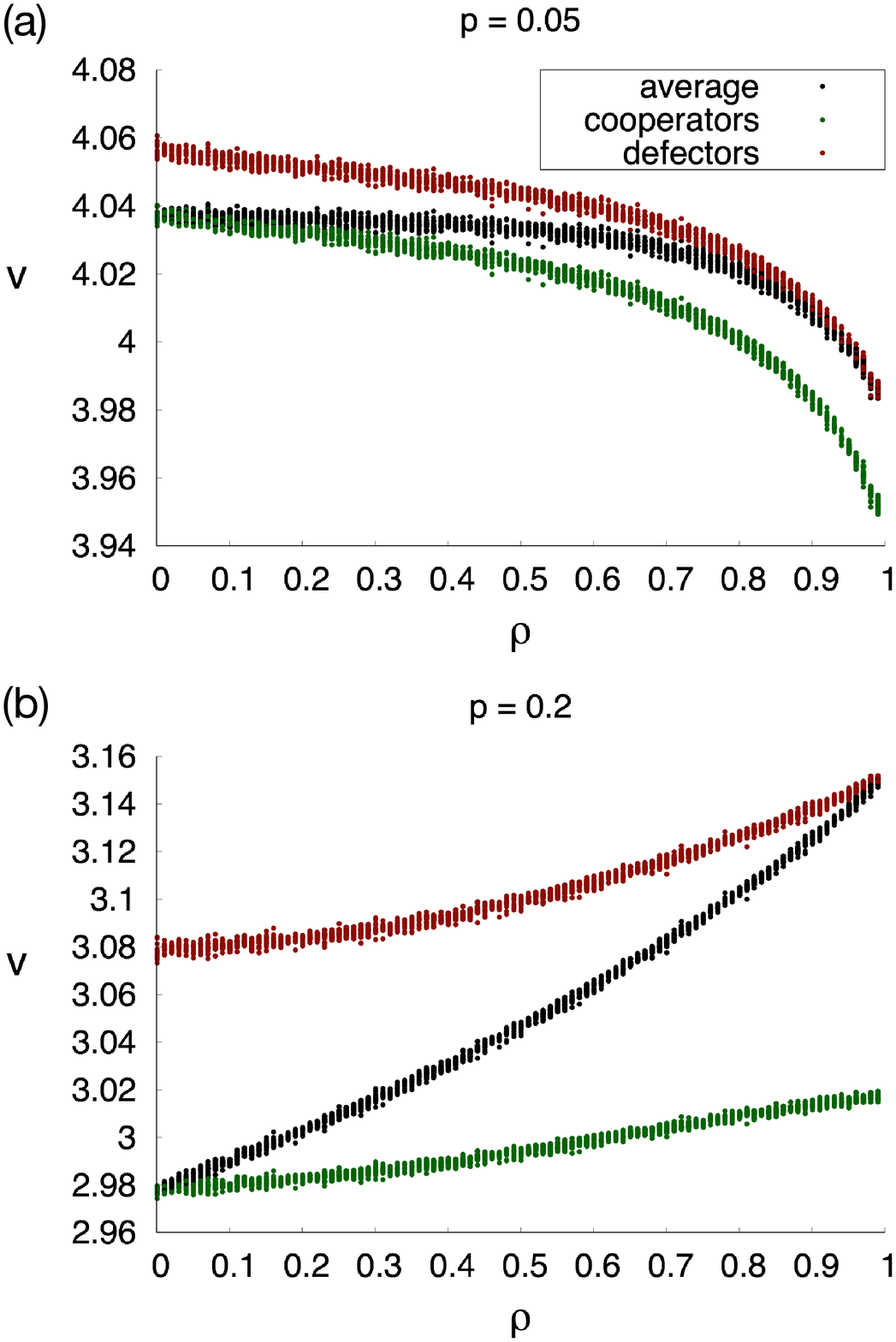}
\caption{ Average velocity as a function of the fraction of defectors. The graph shows simulation data for the average payoff of the defectors (red), cooperators (green), and the average population payoff (black). In (a), where the probability of deceleration is small, the payoff of the defectors is always higher, but the social optima is achieved if everyone cooperates, indicating that defection is a Nash equilibrium \cite{nash1950equilibrium,sigmund1999evolutionary,osborne1994course}. In (b), where the probability of deceleration is higher, the individual self-interest and the social optima are aligned: it is always better for the individual to defect and the social optimum is achieved if everyone defects. The parameters are $c=0.18$ for both graphs and $p=0.05$ in (a) and $p=0.2$ in (b).}
\label{mixed-velocities}
\end{figure}

To characterize the emergence of the social dilemma, we can vary the fraction of the types in the population and measure the variations of their payoffs \cite{hauert2006}. Figure \ref{mixed-velocities} shows the average velocity as a function of the density of defectors. The payoff of the defectors is always greater than that of the cooperators, independently of the concentration of defectors. If the probability of random deceleration is small, the average payoff of the population decreases as the number of defectors increases. Therefore, the population is better off if all individuals cooperate. This scenario characterizes a social dilemma, where the individual's willingness to maximize his own payoff drives the population to a state that is worse than a state where everyone cooperates. This social dilemma is of the same kind that is present in the tragedy of commons and in the prisoner's dilemma game. On the other side, for higher values of the probability of deceleration, the social optimum and the individual self-interest are aligned:  both the defectors' payoff and the average population payoff increases if the fraction of defectors increases. In this scenario, there is no social dilemma: it is the best, both for the individual and for the population, to adopt defection.

\begin{figure}[tbh]
\includegraphics[scale=0.9]{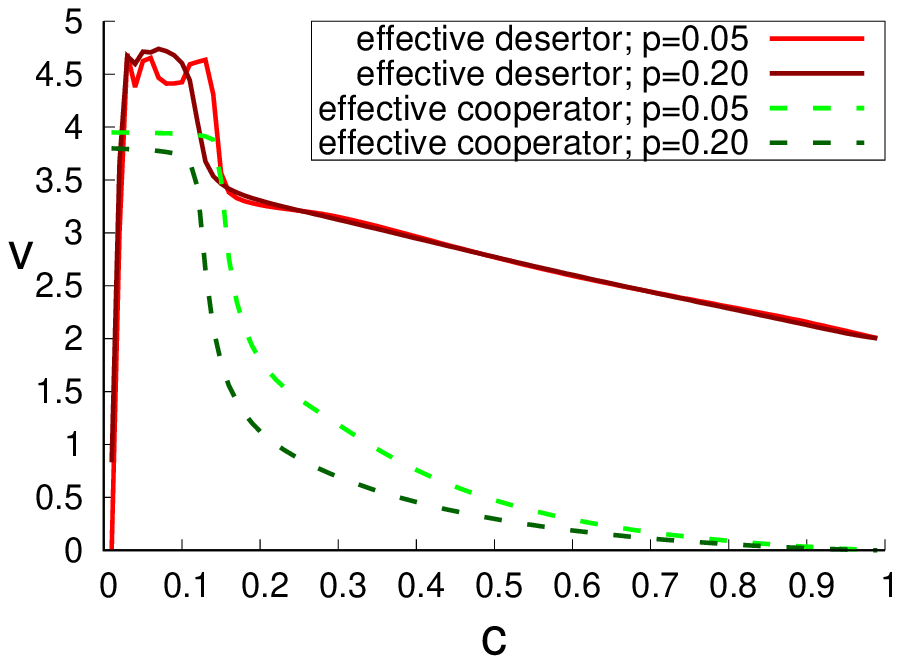}
\caption{Average velocity in the ALLD population formed after aggregation the defectors in terms of their effective behaviour. The defectors that overtake successfully (effective defectors) are shown in full lines and the defectors that fail to overtake (effective cooperators) in dashed lines. Notice that only vehicles that attempt to overtake are included in the averaging. The average velocity of the effective defectors is always higher than that of the effective cooperators. At low concentrations, the velocity of the effective defectors is erratic due to the small number of successful overtaking. Notice that at very high concentrations, c=0.99, the mean velocity of effective defectors is the smaller speed possible allowing overtaking ($v=2$).}
\label{velocities_comparison}
\end{figure}

The difference between the average payoff of the cooperators and defectors is small because the overtaking does not happen in every attempt. Figure \ref{fundamental-diagram}-c shows that, even in the ALLD population, at most only  $2.6\%$ of the vehicles overtake. However, if we measure the average velocity of the defectors that successfully overtakes, the gain is significant, as shown in Fig. \ref{velocities_comparison}.  Therefore, although the effect over the average payoffs is small, the choice of the strategy is relevant to the individual. 

\begin{figure}[tbh]
\includegraphics[scale=0.7]{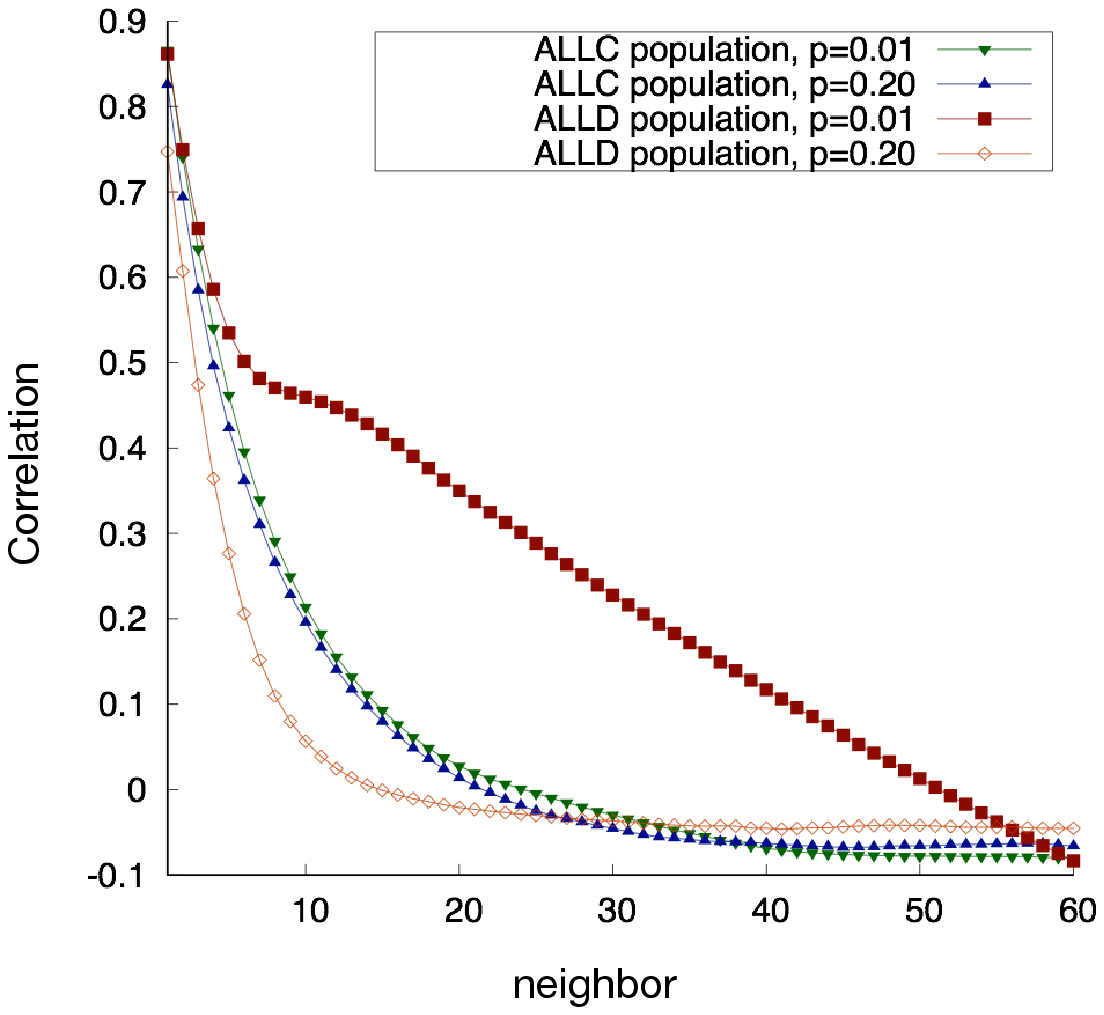}
\caption{Correlation between the velocities of a focal vehicle  and its $j$-th neighbor (x-axis). The graph shows the average taken after considering every car as the focal one in homogeneous populations of cooperators and defectors for low and high $p$ values, all at $c=0.20$. Notice that for the ALLC population the correlation has  a weak dependence on $p$. On the other side, in ALLD population the correlation is relatively strong for $p=0.01$ and weak for $p=0.20$.}
\label{Correlations}
\end{figure}

To have a clearer picture of the non-trivial dependence of the individual payoffs on the random deceleration probability, we analyze the interactions among the vehicles. The interactions take place only when cars have to reduce their velocities due to the short headway distances. Also, the interactions are asymmetrical, that is, the car in the front affects only the car behind. Although the interactions are pairwise, there can be a cascade of braking events affecting many vehicles behind. To quantify the effect of the cascade of braking events, we  measured the correlation of the velocities of the vehicles, as shown in Fig.  \ref{Correlations}. In the ALLC population, the harmful effects of the interactions  persist for many vehicles, but it is almost independent of the random deceleration probability. On the other hand, in the ALLD population, it is strongly dependent on $p$. The correlation is higher at low $p$ values, that is, the overtaking increases the size of the interacting groups and the harmful effects of overtaking spread to the system, with  many vehicles being hindered by the actions of the defector. The correlation drops significantly for high $p$ values, that is, the harmful effects of the interactions are relatively localized and the costs of overtaking are paid by the nearby neighbors of the overtaking defector. Hence, for high $p$ values, the harmful effect of the overtaking does not escalate and the social dilemma does not emerge.

\subsection{Discontinuities at $\rho=1$ and $p=0$ }

The simulation results indicate that the flux, as a function of $c$, $p$ and $\rho$ (concentration of vehicles, probability of random deceleration, and fraction of defectors, respectively), is a continuous function except along the line defined by $p=0$,  $\rho=1$ and $c_{1}<c<c_{3}$. This discontinuity is important because it is related to the problem of invasion of a mutant strategy. In this section, we discuss this discontinuity.

Let us first fix $\rho=1$ so that $J=J(c,p)$. In the limit $p\rightarrow 0$, the behavior of the ALLD population is very similar to that of the deterministic model, which is characterized by the 2-cycle overtaking pattern. Recall that the 2-cycle produces the forward-moving pattern shown in the detail of  Fig. \ref{figp0}. This pattern, however, is very unstable. The perturbations introduced by the random deceleration breake the overtaking pattern and the system behaves like an ALLC population, as overtaking is no longer possible due to the great symmetry produced by the overtaking pattern. Nevertheless, the overtaking pattern reemerges and may take over the population again as a consequence of $p\ne0$. Consequently, the system oscillates between both patterns. Let $J_{C}$ be the characteristic flux of the ALLC population and $J_{D}$ that of ALLD population under this scenario. Let $\tau_{C}$ and $\tau_{D}$ be the characteristic times spent in each state. The flux can be approximated by:
\[
J(c,p)=\frac{\tau_{D}J_{D}+\tau_{C}J_{C}}{\tau_{D}+\tau_{C}}.
\]
The average time spent in the state that exhibits the 2-cycle overtaking pattern is roughly given by $\tau_{D}\approx 1/(Np)$, because the pattern breaks as soon as the overtaking vehicle in the 2-cycle is prevented to overtake  due to the random deceleration. On the other side, the overtaking pattern can emerge again if two events take place. First, one vehicle inside the dense local patch is subjected to the random deceleration, creating a defect inside the patch (composed of a vehicle with velocity $v=1$) and, second, this defect is again subjected to the random deceleration, producing a particle with velocity $v=0$. Thus, $\tau_{D}\approx 1/((N-Lc_{1})p)^{2}$. In the limit $p \rightarrow 0$, $\tau_{D}$ increases faster than $\tau_{C}$ and the system spends more time in the overtaking pattern so that $J(c,p)\rightarrow J_{D}$. Therefore, $J(c,p)$ is continuous at $p=0$.

Let us now fix $p=0$ so that $J=J(c,\rho)$. In this case, any small amount of cooperator in the population will eventually breake the 2-cycle overtaking pattern. Because $p=0$, the 2-cycle overtaking pattern cannot emerge again. Therefore, the system exhibits a discontinuity at $\rho=1$ so that $\lim_{\rho\rightarrow 1}J(c,\rho)= J_{C}$ and $J(c,1)=J_{D}$, where $J_{D}$ and $J_{C}$ are the same as described in the previous paragraph.

Because at $p=0$ we have that $J(c,\rho)= J_{C}$ for $\rho \neq 1$, the flux difference between the ALLC and the ALLD populations is equal to zero, except at $\rho=1$. Hence, the social dilemma that emerges in the deterministic case is just a weaker version of the dilemma.

\subsection{More realistic single-lane algorithms}

To test the robustness of our findings we analyzed a more realistic model in single-lane that incorporates anticipation in order to prevent strong braking events \cite{realistic}. In the original model, the velocity of the leading car is anticipated, allowing smother deceleration in function of the lack of  available space. At large distances, the drivers move as fast as possible and, at intermediate distances, the drivers respond to the brake lights. Finally, at small distances, the drivers adjust their velocity to drive as safe as possible. In particular, the value of the random deceleration $p$ is determined by a set of conditions: if the vehicle is at rest, then $p=p_0$; if the next vehicle has its brake lights on and is inside the interaction radio, then $p=p_b$; in any other case $p=p_d$. If the vehicle decelerates, or $p=p_b$, its brake lights turn on. 

We implemented the overtaking strategy similarly to what we did to the NaSch model. The defector can overtake as many vehicles as possible and, when overtaking is possible, the driver ignores the brake lights of the next vehicle, accelerates if the velocity limit allows, and is subjected to $p=p_d$. The last overtaken vehicle  has its brake light turned on.

\begin{figure}
    \centering
    \includegraphics[scale=0.75]{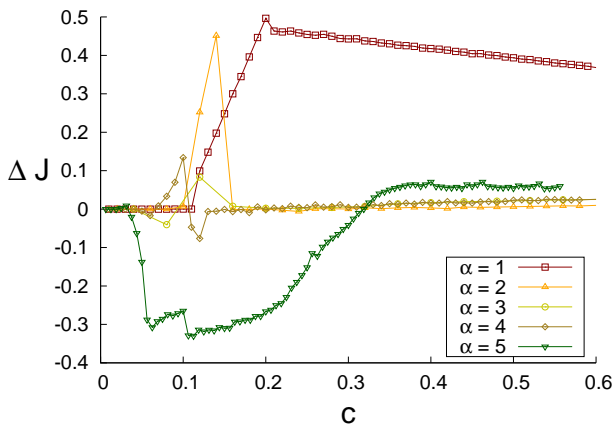}
    \caption{ The difference between the fluxes of ALLD and ALLC population in the $\alpha$-model. The advantage of the overtaking strategy decreases sharply as $\alpha$ increases. The dilemma appears because of the asymmetry between the acceleration (slow process) and deceleration (fast process) in braking events caused by the lack of space.}
    \label{Reality}
\end{figure}

The overreaction of the vehicles that are overtaken is important in our minimal model. Hence, it should also play an important role in the more realistic model. One way to control this effect is to change the scale of parameters so that the size of the acceleration steps are smaller in comparison to the abrupt deceleration caused by the overtaking. More specifically, the maximum velocity $v_{max}$, the size of the vehicle $s$, the security gap $b$,  and the length of the highway $L$ are all increased by the same factor $\alpha$ (we call it the $\alpha$-model). As the time step is unchanged and we have yet to present the results in physical unities, a larger size of the highway under this transformation implies  finer acceleration steps. 
Figure \ref{Reality} shows the flux difference between the ALLC and the ALLD population in the $\alpha-$model. The social dilemma in the more realistic model emerges if $\alpha$ is sufficiently high because it becomes harder to recover the speed after the abrupt deceleration caused by the overtaking. The overreactions occur immediately after the overtaking, causing strong correlations in the velocities of the vehicles. 

We measured the ratio of frustrated overtaking (when the overtaking is initiated but could not be completed), continuous overtaking (when the overtaking vehicle does not decelerate to overtake), and accelerating overtaking (when the overtaking vehicle have to decelerate but overtakes after some time). 

It turns out that at low global density, $31\%$ were frustrated, $6\%$ were accelerating and $62\%$  were continuous. On the other hand, at high global density these ratios change to $93\%$, $2\%$ and $5\%$, respectively. These measurements and comparison with the experimental data suggests that the drivers in the model are very aggressive (in \cite{Mocsari_1,passing1,passing2,impatient} the ratio of continuous overtaking is taken as a indirect measurement of aggressiveness among drivers), but not as aggressive as drivers studied in undivided roads in some parts of India \cite{passing2}.

\section{Discussion and conclusion}
\label{conclusions}

The NaSch model contains the key ingredients to simulate basic features of real vehicular traffic. We introduced overtaking as an action where the driver must analyze if there is space available and if its velocity is sufficient to make the manoeuvre. In more realistic scenarios, the driver would have to analyze the traffic in the contiguous lane with opposite traffic, but we do not add this factor to our model. We started with the minimal NaSch model because of its simplicity, which allows an analytical treatment of the deterministic case. As the analysis of a more realistic model showed, as long as there is an abrupt arrival of the defector in front of an overtaken car, overtaking can create a social dilemma. Hence, the social dilemma may appear in more realistic single-lane traffic models if there is an asymmetry between acceleration, as a slow process, and the deceleration as a response to the lack of space available, as we have shown. Notice that the defector overtakes precisely to avoid this deceleration.

The effect of overtaking depends strongly on the probability of random deceleration.  The random deceleration causes the spontaneous formation of jams in the NaSch model and is one of the factors that control the emergence of the social dilemma when overtaking is allowed. The social dilemma emerges if the probability of random deceleration is small and it is caused by the strong correlations between the vehicles. Interestingly, the random deceleration can be interpreted as a random intrinsic response of the driver to the imperfections of the road \cite{bouadi2016effect}. Under this interpretation, we can say that, if the overall condition of the road is good, the population harvests  better results if all cooperate, being more patient and complying with the ordered flow of vehicles. 

In typical examples of cooperation \cite{Fletcher2009}, the cooperator is the active agent who provides a benefit to the recipient at a cost to himself. The defector is the passive recipient, who free-ride on the cooperative behavior of others. In our model, the defectors are the active agents, because their actions provide benefits to themselves and inflict costs to the population, whereas the cooperators are the passive agents that suffer the costs. Because the interactions may involve more than two individuals, it is not possible to write down a simple payoff matrix as in the two-player prisoner's dilemma. Nevertheless, the nature of the game played by the drivers in our model is clearly one that exhibits a social dilemma, as shown in Fig. \ref{mixed-velocities} and \ref{velocities_comparison}. 
 
The identification of social dilemmas in vehicular traffic systems is important because minimizing the driver's time spent on the road (individual payoff) may not align with the goals of traffic engineering, as the maximization of the flux of vehicles (social payoff at $c$ constant).  Our study shows that, under certain circumstances, overtaking is good because it increases both the road capacity and minimizes the driver's mean time spent on the road. However, when these conditions are not met, overtaking is harmful to all. In this case, one may have to identify mechanisms that promote cooperation. 

Vehicular traffic is a social phenomenon. The high population density in the big cities creates complex social interactions which can certainly be improved if the individuals take cooperation seriously.  As shown here, cooperation is again playing a major role in a basic part of human life: moving from here to there.

\begin{acknowledgements}
The authors thank CNPq, CAPES and FAPEMIG, Brazilian agencies.
\end{acknowledgements}

\appendix*
\section{THE ALGORITHM}

The update starts with the choice of the first car to have its velocity updated. To avoid artificial effects caused by the choice of the first car, in every step we find the car with the largest velocity and we pick the car that comes behind it to be the first to update the velocity. In this way, because  we know for sure  that the first chosen car will not overtake, we don't need any information about the car in front, which is needed in our algorithm. Notice that all information is available for the other cars .  Interestingly, if we had  initiated the update with a randomly chosen vehicle and force this car to behave as cooperator, such small perturbation would destroy the patterns formed in the deterministic case ($p=0$), making the system to have a NaSch-like dynamics. It is similar to the phenomenon observed at the discontinuity at $\rho=1$.

After the first car is chosen, the following update rules are applied. Let $x_n$ and $v_n$ be the position and the velocity of the n-th vehicle. If the driver is a cooperator, the NaSch algorithm is implemented. If the driver is a defector, the following steps are implemented. In the first step, the vehicle accelerates one unity. In the second step, the driver  evaluates the available space available, which is given by $G_{n}=x_{n+1}-x_{n}-1$. If $v_{n}\le G_{n}+v_{n+1}$, then the driver follows the NaSch algorithm, manifesting a cooperator-like behavior. On the other hand, if $v_{n} > G_{n}+v_{n+1}$, then the vehicle is subjected to the random deceleration and, if there is a $j$ such that \begin{equation}\label{fundamentalInequality}
   x_{n+j+1}>x_{n}+v_{n}>x_{n+j}+v_{n+j},
\end{equation} 
the overtaking is successful. The left inequality means that the driver follows the NaSch algorithm, interacting with the $(n+j+1)$-th vehicle (ignoring the others), which is a characteristic of a non-one-dimensional problem. The second inequality means that the defector will not force the overtaken vehicles to decelerate at the moment of the overtaking, which can be regarded as a security condition.  The vehicles may overtake as many vehicles as possible.  As long as $j>1$ and at least one of the inequalities in \ref{fundamentalInequality} is not satisfied, we make $v_{n}=x_{n+j}-x_{n}-1$,  $j=j-1$, and try again, until both inequalities in \ref{fundamentalInequality} are satisfied or $j=0$. If all overtaking attempts are failed,  we are back to the original NaSch algorithm and the driver behaves as a cooperator. The general scheme of the algorithm is the following:

\vspace{0.5cm}
\noindent step 1: \hspace{0.cm} $v_n =\min(v_{n}+1,v_{max})$ FOR ALL $n$; 

\hspace{-.4cm}DO n=M,M-N

\noindent step 2': \hspace{-.2cm} IF$[v'_n \le G_{n}+v_{n+1})]$THEN 

\hspace{1.2cm} $v'_n  =\min(v_n,G_{n})$

\hspace{1.2cm} $v'_n  =\max(v_n-1,0)$, with probability p; 

\hspace{1.2cm} CYCLE

\noindent  \hspace{1.2cm} END IF

\noindent step 2: \hspace{0.cm} IF$[x_{n+j+1}>(v'_n + x_n)>(v'_{n+j} + x_{n+j})]$ 

\hspace{1.0cm} for some $j > 0$; 

       \hspace{1.2cm} $v'_n  =\max(v_n-1,0)$, with probability p; 
      
        \hspace{1.2cm}IF[$(v'_n+x_n)==(v'_{n+j}+x_{n+j})$]  
       
\hspace{1.7cm} $v'_n =  x_{n+j} - x_{n} - 1$;

\hspace{1.6cm} $j=j-1$;

\hspace{1.5cm} GO TO STEP 2';

\hspace{1.2cm}END IF
       
       \hspace{1.2cm}CYCLE  

\hspace{.9cm}  ELSE 

\hspace{1.3cm} $v'_n =  x_{n+j} - x_{n} - 1$;

\hspace{1.3cm} $j=j-1$;

\hspace{1.2cm} GO TO STEP 2';

\hspace{1.0cm}  END IF

\hspace{-.4cm}END DO 

\hspace{-.5cm} $\vec{x'} = \vec{x} + \vec{v'}$;

\hspace{-.5cm} REORDER $\vec{x'}$; 

\vspace{0.5cm}



%

\end{document}